\documentclass[superscriptaddress,aps,amsmath,amssymb,floatfix,reprint,nobalancelastpage,raggedbottom]{revtex4-1}
\usepackage{graphicx}
\usepackage[colorlinks=true,citecolor=blue,linkcolor=blue,urlcolor=blue]{hyperref}
\usepackage{amsmath}
\usepackage{dcolumn}
\usepackage{xcolor}
\usepackage{fancyhdr}
\usepackage{lipsum}
\usepackage{soul}

\usepackage{footnote}

\makeatletter 
\renewcommand{\fnum@figure}{\textbf{FIG.~\thefigure}}
\makeatother

\makeatletter
\def\bbordermatrix#1{\begingroup \m@th
  \@tempdima 4.75\p@
  \setbox\z@\vbox{%
    \def\cr{\crcr\noalign{\kern2\p@\global\let\cr\endline}}%
    \ialign{$##$\hfil\kern2\p@\kern\@tempdima&\thinspace\hfil$##$\hfil
      &&\quad\hfil$##$\hfil\crcr
      \omit\strut\hfil\crcr\noalign{\kern-\baselineskip}%
      #1\crcr\omit\strut\cr}}%
  \setbox\tw@\vbox{\unvcopy\z@\global\setbox\@ne\lastbox}%
  \setbox\tw@\hbox{\unhbox\@ne\unskip\global\setbox\@ne\lastbox}%
  \setbox\tw@\hbox{$\kern\wd\@ne\kern-\@tempdima\left[\kern-\wd\@ne
    \global\setbox\@ne\vbox{\box\@ne\kern2\p@}%
    \vcenter{\kern-\ht\@ne\unvbox\z@\kern-\baselineskip}\,\right]$}%
  \null\;\vbox{\kern\ht\@ne\box\tw@}\endgroup}
\makeatother

\begin{document}
\title{Low Barrier Nanomagnets as p-bits for Spin Logic}
\author{Rafatul Faria}
\email{rfaria@purdue.edu}      
\affiliation{School of Electrical and Computer Engineering, Purdue University, IN, 47907}
\author{Kerem Yunus Camsari}
\affiliation{School of Electrical and Computer Engineering, Purdue University, IN, 47907}
\author{Supriyo Datta}
\email{datta@purdue.edu}      
\affiliation{School of Electrical and Computer Engineering, Purdue University, IN, 47907}
\date{\today}

\begin{abstract}

It has recently been shown that a suitably interconnected network of tunable telegraphic noise generators or ``p-bits'' can be used to perform even precise arithmetic functions like a 32-bit adder. In this paper we use simulations based on the stochastic Landau-Lifshitz-Gilbert (sLLG) equation to demonstrate that similar impressive functions can be performed using unstable nanomagnets with energy barriers as low as a fraction of a $\rm kT$. This is surprising since the magnetization of low barrier nanomagnets is not telegraphic with discrete values of $\pm 1$. Rather it fluctuates randomly among all values between $-$1 and +1, and the output magnets are read with a thresholding device that translates all positive values to 1 and all negative values to zero. We present sLLG-based simulations demonstrating the operation of a 32-bit adder with a network of several hundred nanomagnets, exhibiting a remarkably precise correlation: The input magnets \{A\} and \{B\} as well as the output magnets \{S\} all fluctuate randomly and yet the quantity A+B$-$S is sharply peaked around zero! If we fix \{A\} and \{B\}, the sum magnets \{S\} rapidly converge to a unique state with S=A+B so that the system acts as an adder. But unlike standard adders, the operation is invertible. If we fix \{S\} and \{B\}, the remaining magnets \{A\} converge to the difference A=S$-$B. These examples emphasize a new direction for the field of nanomagnetics away from stable high barrier magnets towards stochastic low barrier magnets which not only operate with lower currents, but are also more promising for continued downscaling. \textbf{Index Terms:} Spintronic memory and logic, nanomagnetics, Landau-Lifshitz-Gilbert equation, arithmetic functions.
\end{abstract}
\maketitle

\section{Introduction}
The developments in spintronics and nanomagnetics are having enormous influence on the field of storage and memory devices and it has been shown that the WRITE (W) and READ (R) elements can also be integrated into units that implement Boolean as well as non-Boolean logic  \cite{bromberg2014experimental,datta2012non,diep2014spin,nikonov2015benchmarking,manipatruni2015spin,sengupta2016proposal,Pan2016,mankalale2016stem}. These applications, however, usually make use of stable magnets with energy barriers $\sim 40 \ \rm kT$ which require relatively large currents for their operation. The critical spin current needed to switch a magnet with a thermal energy barrier of $\Delta = H_K M_s V/2$ is given by \cite{sun2000spin}
\vspace{-8pt}
\begin{equation}
I_c= I_{c0} \frac{\Delta}{kT} \quad  \quad I_{c0} = \frac{4q \alpha}{\hbar}kT \bigg(1+ f_I \ \frac{ H_d}{2 H_K}\bigg)
\label{Ic0}
\end{equation}
\noindent where $q$ is the electronic charge, $M_s$ is the saturation magnetization, $H_K$ is the anisotropy field, $H_d$ is the demagnetization field, V is the volume, $\alpha$ is the Gilbert damping coefficient and the factor $f_I$ is equal to zero for perpendicular anisotropy magnets (PMA) and one for inplane anisotropy magnets (IMA). With $\Delta \sim 40 \ \rm kT$ and $\alpha \sim 0.01$, the critical switching spin current for a PMA magnet is $ 4q \alpha \Delta / \hbar \approx 10 \ \mu A.$ Magnets with lower barriers could operate with lower currents but their application in conventional memory or logic is severely limited due to their stochastic nature. However, their possible use in unconventional applications has been discussed both theoretically and experimentally \cite{locatelli2014noise,bapna2016magnetostatic, piotrowski2014magnetic, choi2014magnetic, mizrahi2016controlling, srinivasan2016magnetic, vincent2015spin, khasanvis2015self,locatelli2015spintronic}. The implementation of logic operations based on an ensemble average over stable nanomagnets has been explored in \cite{behin-aein_building_2016,bai2016stochastic,  shim2016ising}  while \cite{sutton2016intrinsic} describes an approach to the traveling salesman problem based on a time average over unstable nanomagnets that cycle through millions of collective correlated states potentially at GHz rates. Note that for such nanomagnets ($\Delta \ll 25 \rm \ kT$ \cite{lopez2002transition}), the Arrhenius model that predicts a telegraphic change between two magnetizations is no longer applicable, and the magnetization becomes a continuous variable.  The present paper describes the application of the latter approach (time average) to implement precise Boolean logic operations like a 32-bit adder that provides the sum S for given inputs A and B. Remarkably the adder also evaluates the inverse function, cycling through all combinations of A and B that add up to a given sum S.
\begin{figure}[h!]
\includegraphics[width=0.85\linewidth]{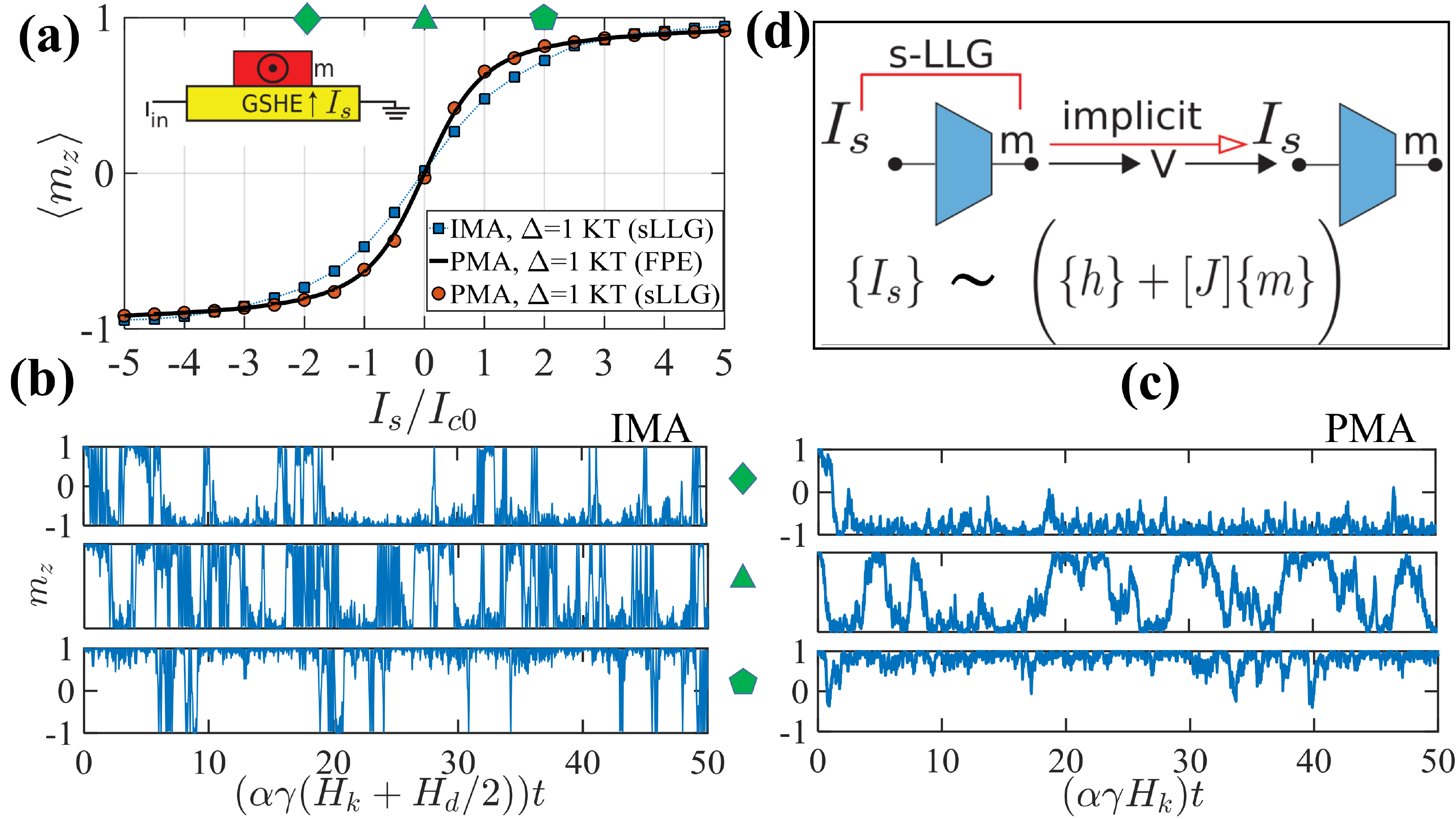}
\caption{\textbf{Low-barrier stochastic Nanomagnet as a p-bit:} (a) Time-averaged magnetization of low barrier IMA and PMA magnets ($\Delta= 1 \ \rm kT,  H_K= 60 \ \rm mT, \alpha=0.01, H_d = 1.5 \ T \ for \ IMA$) as a function of the bias spin current which is normalized to $I_{c0}$ (Eq. \ref{Ic0}). Average magnetization of PMA magnets obtained from sLLG which agrees well with the analytical solution from the FPE, Eq. \ref{Pm}. Inset shows a physical structure using a giant spin Hall effect (GSHE) material that could be used to convert a charge current into a spin current with the correct polarization to bias an IMA. (b) The magnetization m(t) for IMA as a function of time for three  different bias currents obtained from a numerical solution of sLLG equation. (c) Same plot for PMA with the same barrier height. Note that the fluctuations are much faster and more telegraphic for IMA than for PMA. (d) A connection scheme for two p-bits is shown where the magnetization of a p-bit is implicitly converted into the bias current/voltage for the next p-bit (Eq. \ref{Ik}). A possible hardware implementation to turn the magnetization m into a voltage V,  could combine a GSHE layer with MTJs as in \cite{datta2012non},  replacing the stable write magnets by low barrier nanomagnets that are discussed here.}
\label{fi:fig1} 
\end{figure}
We have recently shown \cite{camsari2016stochastic} that a suitably interconnected network of tunable telegraphic noise generators or telegraphic ``p-bits'' can be used to perform even precise arithmetic functions like a 32-bit adder. However, it is not clear whether such p-bits can be implemented with real physical systems, especially if the noise in these systems are not telegraphic but continuous. The objective of this paper is to demonstrate that p-bits can be implemented using unstable nanomagnets with energy barriers as low as a fraction of a kT, \textit{even though their magnetization is not telegraphic} and fluctuate  among all values from $-$1 to +1. We assume that the magnets can be read with a thresholding device that translates all positive values to +1 and all negative values to zero. But this thresholding is applied \textit {only} to the output nodes when we need to read a magnet at the end of an operation and not to the internal nodes or during device operation.

\begin{figure}[t!]
\includegraphics[width=0.85\linewidth]{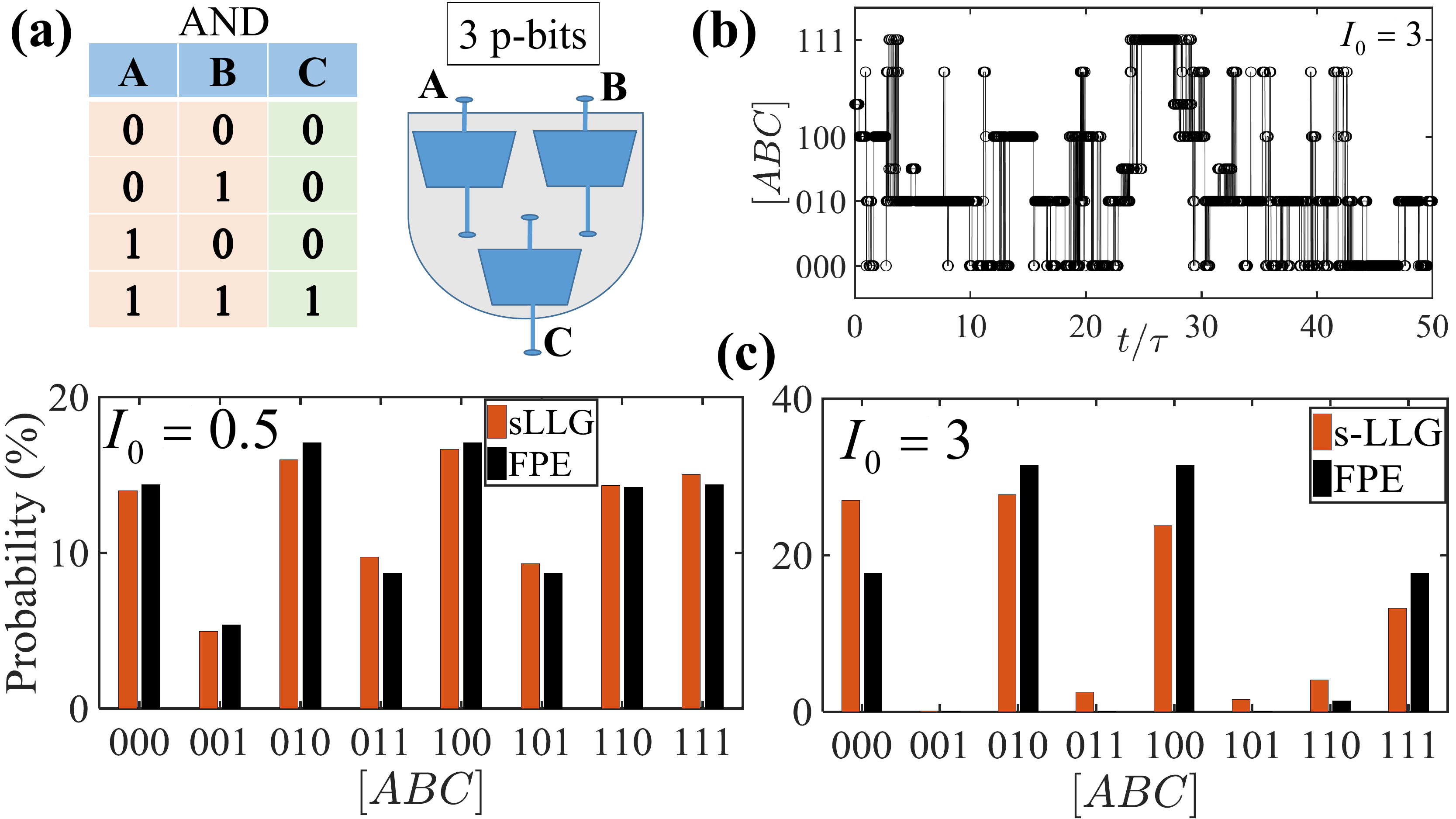}
\caption{\textbf{Implementation of a basic boolean element (AND) using p-bits:} (a) The truth table for AND is shown along with a schematic for the network of three p-bits used to perform the operation. The p-bits are connected symmetrically with $J_{ij}=J_{ji}$. (b) The decimal value of each configuration of the input-output nodes at each time step (normalized by the factor $\tau=(\alpha \gamma (H_k+H_d/2))^{-1}$) is calculated according to $A\times2^2+B\times2^1+C\times2^0$ where A, B and C are thresholded to obtain binary values (0,1) at the read out. (c) Histograms of the different configurations of the p-bits are shown for a weaker $(I_0=0.5)$ and stronger $(I_0=3)$ correlation strength. Note the close match between the numerical values obtained from the sLLG equation with the probabilities obtained analytically from the FPE result in Eq.~\ref{Ps} which is related to the Boltzmann law, especially for $I_0$=0.5. For higher values of $I_0$ the numerical results tend to be stuck in metastable states requiring longer simulation times to converge to the steady-state FPE result.}
\label{fi:fig2}
\end{figure}

 We start in \textit{Section 2} by showing that low barrier magnets, both PMA and IMA, exhibit the key property of p-bits, namely that they act as electrically tunable random number generators (RNG). Their magnetization $m(t)$ fluctuates randomly in time, and the time-averaged $\langle{m}\rangle$ can be tuned from $-$1 to +1 with a spin current. IMA magnets require a larger current to tune, but this is offset by a more rapid fluctuation rate, allowing a faster evaluation of the time average, and hence faster operation (Fig.~\ref{fi:fig1}). Note also that the PMA magnetization is relatively continuous compared to IMA magnetization which is more telegraphic in nature.

To harness either for logic applications, they have to be interconnected such that the spin current $I_{sk}$ driving magnet `k' has to be derived from the magnetization of other magnets.
\vspace{-5pt}
\begin{equation}
\frac{2I_{sk}}{I_{c0}}= - I_0 \bigg(h_k + \sum_j J_{kj} m_j \bigg)
\label{Ik}
\end{equation}
\noindent where $I_{c0}$ is normalization constant defined as the critical current  (Eq. \ref{Ic0}) for a magnet with a barrier $\Delta = 1 \rm \  kT$ and $I_0$ determines the overall strength of the interconnections. The bias $\{h\}$ and interconnection $[J]$ matrices have to be designed appropriately in implementing specific operations. We will not go into the implementation of these matrices since there are many options requiring careful discussion \cite{diep2014spin},\cite{sengupta2016proposal},\cite{yang2013memristive},\cite{yamaoka2016ising}. We will assume that a network of stochastic nanomagnets (PMA and IMA) has been interconnected according to Eq. \ref{Ik} and simulate their behavior using the stochastic Landau-Lifshitz-Gilbert (sLLG) equation to demonstrate useful functionalities. We assume that the currents specified by Eq.~{\ref{Ik}} are applied to each magnet on a time scale that is much shorter than the magnet dynamics, and new features could arise if delays associated with these interconnections are comparable to magnet dynamics. These issues are beyond the scope of this paper. All numerical examples are presented for IMA with parameters shown in Fig.~\ref{fi:fig1} but similar results are obtained with PMA as well.

In \textit{Section 3} we describe how simple logic gates can be implemented by suitably designing the $\{h\}$ and $[J]$ matrices so that the magnet configurations corresponding to the desired truth table represent `low energy' states where the network spends most of its time according to the Boltzmann law of equilibrium probabilities: $P(\{m\}) \sim \exp (- {E(\{m\})}/{kT})$. Although the use of spin currents does not in general permit us to write an energy functional \cite{PhysRevLett.94.127206}, for symmetrically interconnected PMA magnets we can use a functional of the form \cite{pinna2013transmag,sutton2016intrinsic}:
\vspace{-5pt}
\begin{equation}
\footnotesize- \frac{E(\{m\}) }{kT}= \sum_{i} \frac{\Delta_i}{kT} \ m_i^2 + I_0 \bigg( \sum_{i} h_i m_i +  \frac{1}{2} \sum_{i,j} J_{ij} m_i m_j \bigg)
\label{E}
\end{equation}\
\noindent to describe the network of interconnected magnets. This can be seen by noting that from the Boltzmann law and Eq. \ref{E}
\begin{equation*}
\frac{\partial \ ln P}{\partial \ m_k} = 2 \frac{\Delta_k}{kT} m_k + I_0 \bigg( h_k + \frac{1}{2} \sum_{j} (J_{kj} +J_{jk}) m_j \bigg)
\end{equation*}
\noindent so that for a symmetric $[J]$ matrix, from Eq. \ref{Ik} 
\vspace{-5pt}
\begin{equation}
P(m_k) \sim \exp \bigg( \frac{\Delta_k}{kT} m_k^2 - \frac{2I_{sk}}{I_{c0}} m_k \bigg)
\label{Pmk}
\end{equation}
\noindent which is exactly the steady-state condition for magnet `k'  that we would obtain from the Fokker-Planck equation (FPE) (\cite{butler2012switching} Eq.~(3.9)) for PMA. Moreover, our ``empirical'' results show that the energy functional shows good agreement even when magnets have an additional shape anisotropy. Note that even though $I_{c0}$ is size-independent, the distribution of the nanomagnet depends on size through $\Delta$: for higher $\Delta$ magnets, more spin current is required to pin the magnetization. We will refer to Eq. \ref{Pmk} as the FPE probability.

\begin{figure}[t!]
\includegraphics[width=0.85\linewidth]{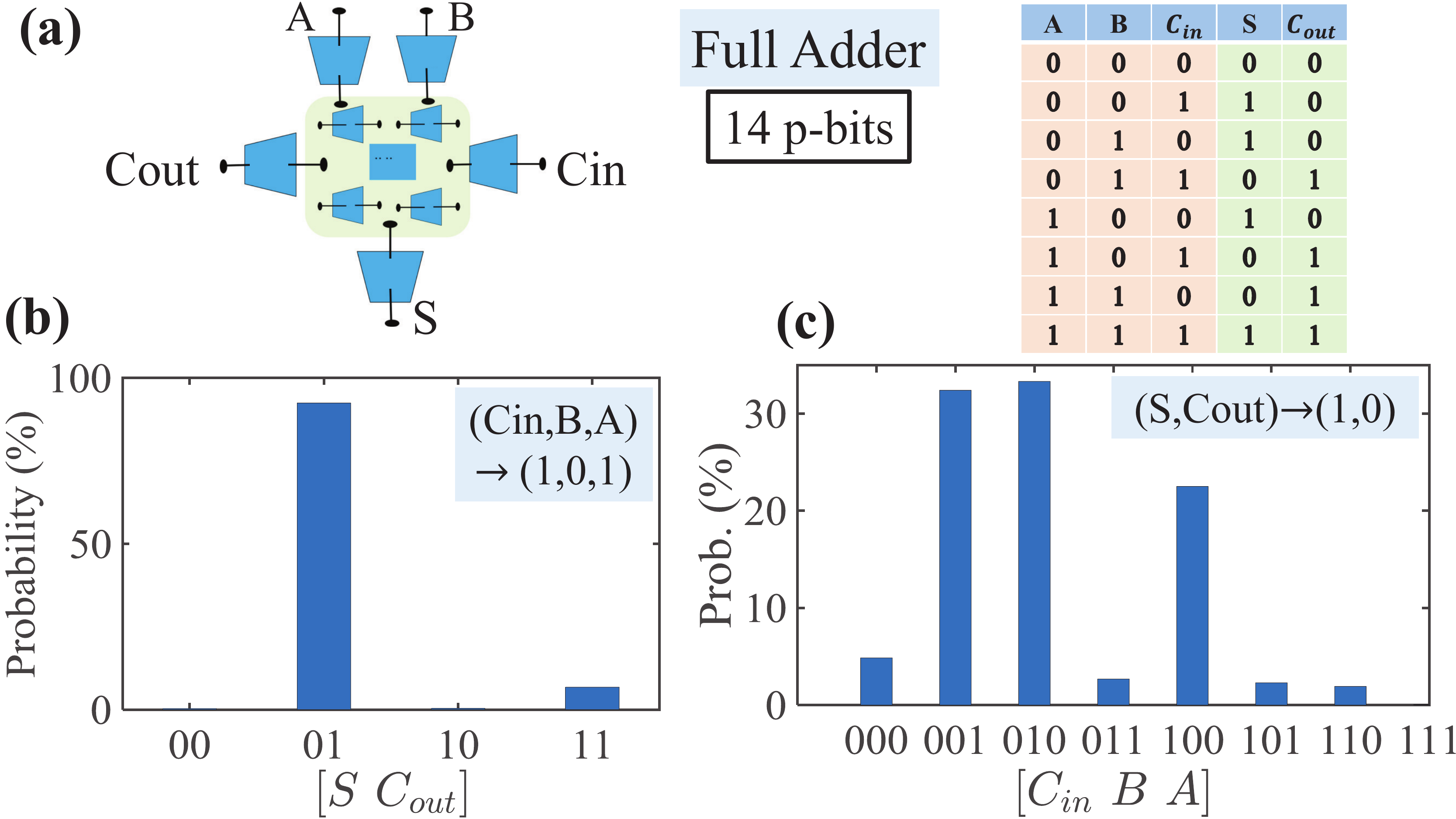}
\caption{\textbf{Full Adder}:  (a) A full adder (truth table shown) implemented by connecting 14 p-bits symmetrically. (b) In forward mode, when the inputs $(A, B, C_{in})$ are clamped, the adder gives the correct output $(S \ and \ C_{out})$. (c) Unlike standard logic, these gates \textit{are invertible}: If the output nodes of the adder are clamped to fixed values, the adder gives all possible input combinations satisfying the output constraint. }
\label{fi:fig3}
\end{figure}
%%%%%%%%%%%%%%%%%%%%%

%%%%%%%%%%%%%%%%%%%%%%
The probability distributions obtained from the numerical solution of the sLLG equation for both PMA and IMA magnets follow the FPE result quite well (Fig.~2). The highest probabilities correspond to the lowest energy states, which correspond to the desired truth table relating the input magnets A and B to the output magnet C. If we force the inputs A and B to specific values by using appropriate values for $h_A$ and $h_B$, C would take on the specific value required by the truth table, just like standard digital gates. But unlike standard gates, these gates are invertible, similar to those discussed in the context of memcomputing \cite{di2016topological}. They can be operated in reverse: if we clamp the output C to a specific value, the inputs A and B will spend most of its time in those configurations $\{AB\}$ that produce that output. We also illustrate this reversible operation with a more complex logic gate, namely a full adder treating it as a Boltzmann machine (BM)  and using the same principle of energy minimization to design the $\{h\}$ and $[J]$ matrices.

\begin{figure}[t!]
\includegraphics[width=0.85\linewidth]{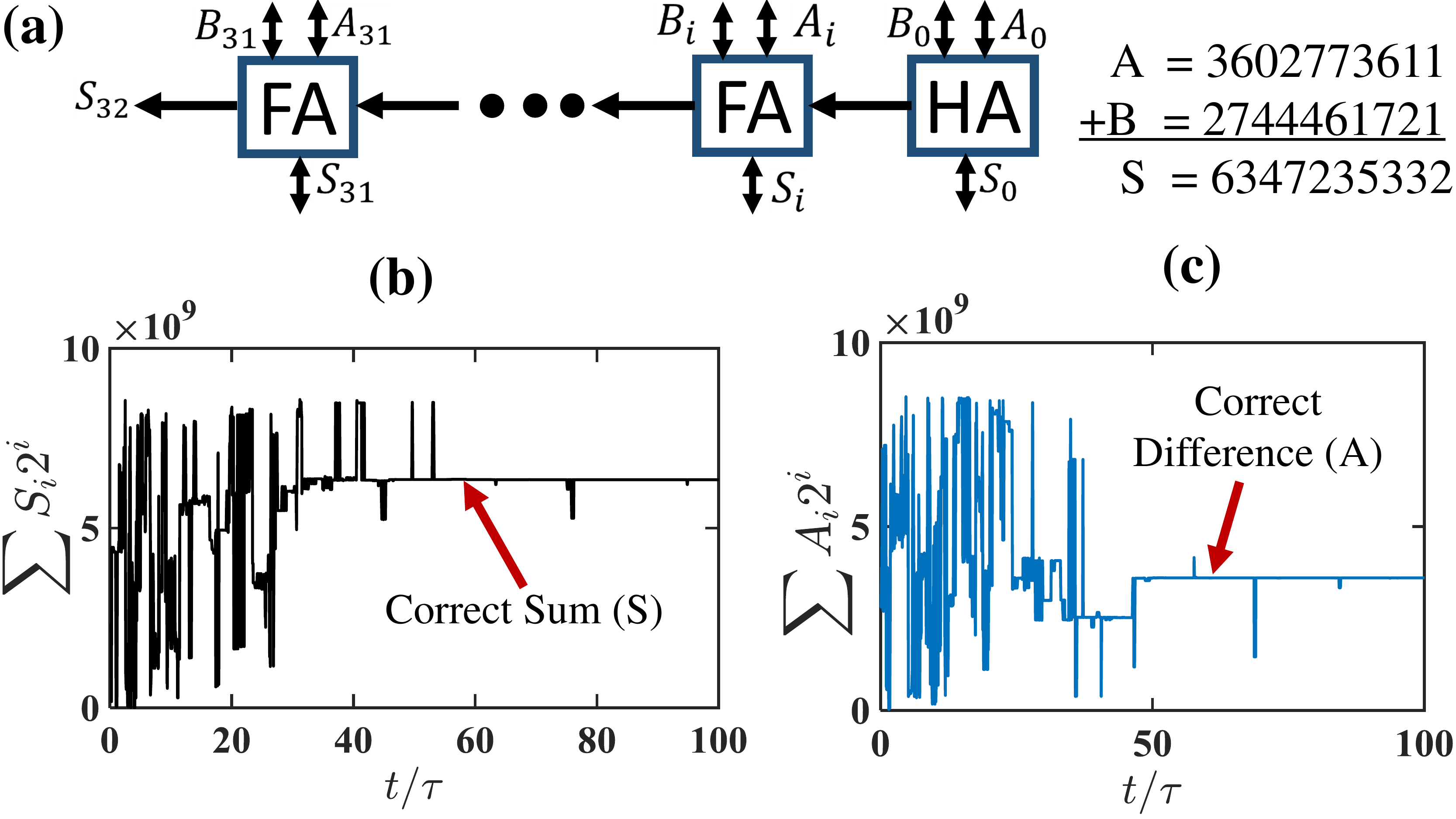}
\caption{\textbf{32-bit Adder/ Subtractor:} (a) Schematic of an adder constructed from 31 full adders (from Fig.~\ref{fi:fig3}) and one half adder (composed of 6 p-bits) with the carry out bit $ C_{out}$ from each adder communicated in a directed fashion to the carry in bit $C_{in}$ of the next adder. (b) Time evolution of the sum $S=\sum_{i}{S_i2^i}$ obtained from the sum bits \{S\} as the coupling strength $I_0$ is ramped up starting from zero. Note that in a time $\sim 60\ \tau$ ($\tau$ is defined in Fig.~2), the sum converges (with occasional jumps) to the correct value which represents one out of $2^{33}\sim8$ billion possibilities. (c) Although the individual adders are connected in a directed fashion through the carry bits, the overall 32-bit adder performs the inverse function as well. If the sum bits \{S\} are clamped along with one set of input (B), the other input converges rapidly to the correct difference (A). }
\label{fi:fig4}
\end{figure}

Finally in \textit{Section 4}  we demonstrate the operation of a 32-bit adder obtained from 31 full adders and one half adder with the output carry from each bit connected to the carry in of the next higher bit through the appropriate element of the overall $[J]$ matrix.  Note that these are unidirectional connections so that the overall $[J]$ matrix is not symmetric, though the $[J]$ matrix for each full adder is symmetric. We show that this network of nearly five hundred nanomagnets exhibits a remarkably precise correlation that provides the exact sum S of any two given inputs, A and B (Fig.4). What is even more remarkable is that if we do not fix either the inputs or the outputs, the quantities A, B and S all fluctuate randomly and yet the quantity A+B$-$S is sharply peaked around zero, so that the network can be used to extract either A, B or S, if the other two are fixed, which is similar to the NP-complete ``subset sum'' problem (Fig.5) \cite{murty1987some, traversa2015polynomial}.

%%%%%%%%%%%%%%%%%%%%%%%%%

%%%%%%%%%%%%%%%%%%%%%%%%%
\vspace{-15pt}

\section{Stochastic nanomagnet model}
\vspace{-12pt}
Fig.1(b,c) shows the time response of the magnetization $m_z$ along the easy axis calculated using the sLLG equation (integrated by Heun's method within the Stratonovich calculus \cite{behin2010proposal}) with $\Delta t = 0.95 \ ps $ for IMA and  $\Delta t = 11.8 \ ps $ for PMA.
%%%%%%%%%%%%%%%%%
\begin{subequations}
\begin{align}
&(1+\alpha^2)\frac{d\hat m_i}{dt} = -|\gamma|{\hat m_i \times \vec{H}_i} - \alpha |\gamma| (\hat m_i \times \hat m_i \times \vec{H}_i)\nonumber \\ &+  \frac{1}{q  N_i}(\hat m_i \times \vec{I}_{Si} \times \hat m_i)  + \left(\frac{\alpha}{q N_i} (\hat m_i \times \vec{I}_{Si})\right)
\label{sLLG}
\end{align}
\end{subequations}

\begin{figure}[t!]
\includegraphics[width=0.99\linewidth, scale=1]{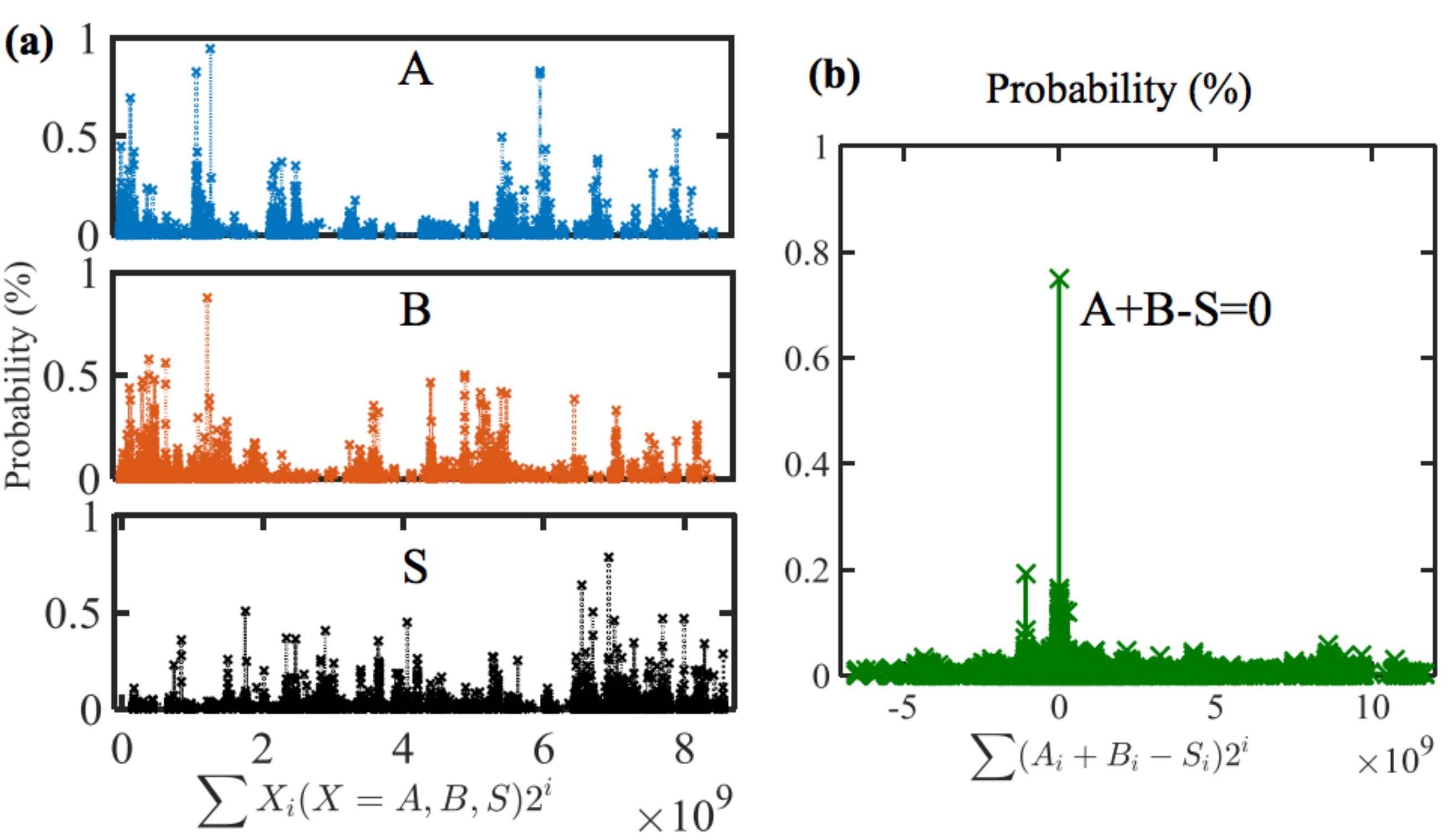}
\caption{\textbf{Correlated Adder}: A remarkable property of the adder (in Fig. \ref{fi:fig4}) is that it works even when the inputs (A,B) and the output (S) are not unique and fluctuate in time amongst many allowed values as shown in (a). Nevertheless, the quantity A+B-S is sharply peaked at zero (b), demonstrating the correlation of hundreds of nanomagnets consistent with the addition function A+B$-$S=0. }
\label{fi:fig5}
\end{figure}

\noindent where $H_i$ is the effective field including the uniaxial and shape anisotropy terms, as well as the thermally fluctuating magnetic field due to three dimensional uncorrelated thermal noise $H_n$ having Gaussian distribution with mean $\langle{H_n}\rangle=0$ and standard deviation $\langle H_n^2\rangle= 2\alpha \rm kT / |\gamma| M_sV$ along each direction \cite{scholz2001micromagnetic, sun2006spin, behin2010proposal, brown1963thermal,vinh2015thesis}, $\gamma$ is the gyromagnetic ratio and $N_i=M_sV/\mu_B$ is the total number of Bohr magnetons comprising the magnet. Our simulations are based on the macrospin approximation, as is common in the literature \cite{lopez2002transition, accioly2016role, adam2009macrospin}. This approximation may not be adequate for larger magnets with multiple domains, but is expected to work better as the magnets are scaled down. The time-averaged magnetization (Fig.~1a) obtained from the sLLG equation for PMA magnets is in good agreement with that obtained analytically by averaging over the  FPE result (Eq.~\ref{Pmk}):
%%%%%%%%%%%%%%%%%
\vspace{-8pt}
\begin{equation}
\langle{m}\rangle = \int_{-1}^{+1} dm \ m \ P(m) \bigg/  \int_{-1}^{+1} dm \ P(m)
\label{Pm}
\end{equation}
\vspace{-28pt}
%%%%%%%%%%%%%%%%%
\section{Basic Boolean Gates}
\vspace{-12pt}
In implementing any given truth table we need the $\{h\}$ and $[J]$ matrices that make the truth table correspond to the lowest energy states of the energy functional given in Eq. \ref{E}. The choice of these matrices is not unique and \cite{biamonte2008nonperturbative} provides a suitable set for AND, OR gates along with many other functions. Fig.~2a shows one possible implementation of an AND gate using a network of three nanomagnets, representing A,B and C.

 The magnetization of the magnets A, B and C fluctuates continuously between $-$1 and +1 and are mapped into the binary values of 0 and 1 by a thresholding operation: all negative values map to zero, while positive values map to +1. The y-axis in Fig.~2b shows the resulting binary number $\{A B C\}$ converted into a single number $A\times 2^2+B\times2^1+C\times2^0$. Note how the values on the y-axis are clustered around 0, 2, 4 and 7 which correspond to the lines of the truth table shown in Fig.~\ref{fi:fig2}a. Occasionally the system jumps to other values but it quickly returns to one of these preferred values. 
 
 This clustering is reflected in the histogram constructed from 678 normalized time steps (Fig.~\ref{fi:fig2}c) which shows peaks around the preferred states defined by the truth table. This agrees well with the probability plot constructed from the FPE result in Eq.~\ref{Pmk} noting that we can label the thresholded states as $ m_i = s_i  m $ where $\quad s_i = \pm 1$ and   $0<m <1$ so that from Eq.~\ref{E}:\\

\footnotesize
\begin{equation*}
E(\{s\},m)\hspace{-3pt} = \hspace{-3pt} \bigg(\sum_{i} \frac{\Delta_i}{kT} +  \frac{1}{2} I_0 \sum_{i,j} J_{ij} s_i s_j\bigg) m^2 + \bigg(I_0 \sum_{i} h_i s_i\bigg) m
\end{equation*}
\begin{equation}
P(\{s\})  \sim \int_{0}^{1}  dm \  \exp \big(-E (\{s\},m)\big)
\label{Ps}
\end{equation}
\normalsize
The peaks corresponding to the preferred states in Fig.~2c do not have equal probability, even at steady state as predicted by Eq.~(\ref{Ps}). This skew is due to the continuous nature of magnetization with small $\Delta$ magnets that affect the thresholded results.

Note that the probabilities are strongly affected by the choice of $I_0$ as we might expect from the exponential dependence of the Boltzmann function. If we use a much smaller value of $I_0$ we obtain a uniform probability across all eight states as we would expect for three uncorrelated magnets. If we use a much larger value of $I_0$ the Boltzmann law predicts all states with equal energy to be equally occupied, but in a numerical simulation, the system tends to get stuck for long periods in one of the preferred states, instead of moving freely among them.

Consider now a full adder having three inputs $A, B, C_{in}$ and two outputs $S, C_{out}$, $S$ being the sum bit, and $C_{in}, C_{out}$ being the incoming carry and the outgoing carry bits. Fig.~3 shows a full adder constructed out of 14 p-bits treating it as a BM with a symmetric J-matrix (\cite{note_j_matrix}) which is obtained by a suitable extension of the principles developed in the context of Hopfield networks (\cite{amit1992modeling}, Eq.~4.20) and extended in \cite{camsari2016stochastic}.  This design not only gives the correct output for a given input, \textit{but also the correct set of inputs for a given output}.

\vspace{-15pt}
\section{32-Bit Adder/Subtractor}
\vspace{-10pt}
Finally we demonstrate the operation of a 32-bit adder obtained from 31 full adders and one half adder with a single directed connection from the $C_{out}$ of one bit to the $C_{in}$ of the next bit, in accordance with the standard design of ripple carry adders (RCA). Here, we treat the RCA as a standalone block without any peripheral read-out circuitry to simply demonstrate how the nanomagnet network can operate as a directed combinational logic unit. If we provide two input numbers A and B, and look at the sum S, which includes all the sum bits along with the carry-out from the last bit, $C_{out}(32)$ we find numerically that the system relaxes to the correct sum with occasional jumps from the correct state. It is really quite surprising that a network of $14\times31+6=440$ nanomagnets fluctuating continuously over the range $-1 < m < +1$ get correlated precisely enough to point to the correct answer out of   $2^{33} \approx 8 \ billion$ possibilities without getting stuck in metastable states \cite{camsari2016stochastic}. Interestingly it also works as a subtractor: if we fix the sum and one of the inputs B, theremaining input gives the correct difference $A=S-B$ (Fig.~4). Even more surprisingly, the overall system seems to act like a BM when all magnets are allowed to fluctuate. Each set of magnets A, B and S fluctuates randomly over a wide range of values. But the quantity A+B$-$S shows a sharp peak around zero (Fig.~5), showing that the interconnected network reflects the desired truth table.
\vspace{-10pt}
\section*{Acknowledgment}
\vspace{-10pt}
The authors gratefully acknowledge many helpful discussions with Behtash Behin-Aein, Vinh Quang Diep, and with Ernesto E. Marinero. This work was supported in part by C-SPIN, one of six centers of STARnet, a Semiconductor Research Corporation program, sponsored by MARCO and DARPA, in part by the Nanoelectronics Research
Initiative through the Institute for Nanoelectronics Discovery and Exploration (INDEX) Center, and in part by the National Science Foundation through the NCN-NEEDS program, contract 1227020-EEC.

\vspace{-10pt}
%\bibliography{PSL}

\begin{thebibliography}{40}
\bibitem{note_j_matrix}
The design of [J] matrices has been discussed in Refs [25, 42] and are assumed not to change during operation.
\end{thebibliography}

%

\end{document}